\documentclass[prl,twocolumn,showpacs,amsmath,amssymb]
{revtex4}
\usepackage{graphicx}
\usepackage{dcolumn}
\usepackage{bm}
\usepackage{txfonts}
\usepackage{mathrsfs}
\usepackage{feynmf}
\usepackage{comment}
\usepackage{float}

\begin{document}
\title{Dicke quantum spin glass of atoms and photons}

\author{Philipp Strack}
\affiliation{Department of Physics, Harvard University, Cambridge MA 02138}
\author{Subir Sachdev}
\affiliation{Department of Physics, Harvard University, Cambridge MA 02138}

\date{\today}

\begin{abstract}
Recent studies of strongly interacting atoms and photons in optical cavities have rekindled interest in the 
Dicke model of atomic qubits coupled to discrete photon cavity modes.
We study the multimode Dicke model with variable atom-photon couplings. We argue that 
a quantum spin glass phase can appear, with a random linear combination of the cavity modes superradiant. 
We compute atomic and photon spectral response functions across this quantum phase transition, both of which 
should be accessible in experiments.  
\end{abstract}


\maketitle

{\textit{Introduction}}.
Ultracold atoms in optical cavities have emerged as attractive new systems for studying 
strongly-interacting quantum many body systems. Photon exchange can mediate long-range interactions
between the atomic degrees of freedom, and this opens up rich possibilities for correlated phases. 
In the celebrated atomic realizations of the superfluid-insulator quantum phase transition \cite{bloch}, 
the light field acts in a secular manner, creating a potential which traps the atoms in an optical lattice;
consequently the atom-atom interactions are only on-site, and this limits the range of possible phases.
In contrast, the seminal recent experiments of Baumann {\em et al.\/} \cite{baumann10,baumann11}, realizing a supersolid phase,
have long-range interactions mediated by active photon exchange \cite{maschler05}. 

Baumann {\em et al.\/} argued that their experiments could be described by the Dicke model, as in the proposal
of Nagy {\em et al.\/} \cite{nagy10}.
The Dicke model couples photons in a single cavity mode uniformly to $N$ atomic two-level
systems (`qubits'). In the limit $N \rightarrow \infty$, this model exhibits a phase transition \cite{hepp73,wang73,emary03,ye11} 
to a ``superradiant'' phase when the atom-photon
coupling is strong enough. In terms of the qubits, the superradiant phase is a `ferromagnet' which spontaneously breaks a 
global Ising symmetry, and so we refer to it as $\text{FM}_{\text{SR}}$. In the experiments by Baumann {\em et al.\/}, 
the superradiance of the cavity photon mode is accompanied by `self-organization' of the atoms into a density wave 
pattern \cite{domokos02,gopa09,gopa10}.

Here we study extensions of the Dicke model to multiple photon cavity modes, and with non-uniform couplings between
the atomic qubits and the photon modes. Spatial mode variations for the single-mode Dicke model were considered in 
Ref.~\onlinecite{larson09}. Multimode Dicke models have been studied earlier \cite{hepp73,EK,thompson,tolkunov07}, 
but were simplified by ignoring
the variations in the atom-photon couplings. We argue here that qualitatively new physics emerges in the multimode case 
when the spatial variation is treated seriously. 
We show that large variations in the atom-photon couplings can give rise to a quantum spin-glass (QSG) phase. 
We will describe quantum-critical dynamics associated with the onset of this spin glass order. 

Dimer {\em et al.} \cite{dimer07} have discussed an experimental realization 
of the Dicke model using internal atomic degrees of freedom, that is, Raman transitions between multiple atomic levels. 
We expect that such schemes can be generalized to a multimode Dicke model that respects a global Ising symmetry, which 
is then {\em spontaneously} broken in the $\text{FM}_{\text{SR}}$ and QSG phase, respectively.
More specific realizations of the multimode Dicke model were described recently by Gopalakrishnan {\em et al.}, in
a paper \cite{gopa11} which appeared while our work was being completed.
The same authors had previously outlined how Bose-Einstein condensates in multimode 
cavities can lead to frustration and glassy behavior \cite{gopa09,gopa10}.
Such experiments on the multimode Dicke model would provide a unique realization
of a quantum spin glass with long-range couplings, and provide a long-awaited testing ground for theories of quantum systems
with strong interactions and disorder. Condensed matter realizations of quantum spin glasses have shorter-range couplings,
and so do not directly map onto the theoretically solvable systems analyzed in the present paper.

Before describing our computations, we point out a key distinction between the transitions involving onset of
$\text{FM}_{\text{SR}}$ versus QSG order. In the single-mode Dicke model,
all the qubits align
in a common direction near the $\text{FM}_{\text{SR}}$ phase, and can therefore be described by a collective spin of length $N/2$ which behaves classically in the limit of large $N$.
Consequently, the dynamics near the phase transition can be described by classical equations of motion \cite{keeling}, and the single-mode Dicke model does not
realize a {\em quantum\/} phase transition in the conventional sense of condensed matter physics. 
In contrast, we will argue here that the onset of QSG order in the multimode Dicke model
has non-trivial quantum fluctuations even in the limit of large $N$, and the critical properties cannot be described by an 
effective classical model. Experimental studies are therefore of great interest.

{\textit{Model}}. The Hamiltonian of the multimode Dicke model is 
\begin{align}
H =\sum_{i=1}^M \omega_i a^\dagger_i a_i 
+
\frac{\Delta}{4} \sum_{\ell=1}^N\sigma_\ell^z
+
\sum_{\ell=1}^{N} \sum_{i=1}^{M}g_{i \ell} \left( a_{i} + a_i^\dagger \right)\sigma^x_\ell\;.
\label{eq:dicke_hamiltonian}
\end{align}
This describes $N$ two-level atomic qubits with level splitting $\Delta/2$ and $M$ photon modes with frequencies $\omega_i$ 
coupled by an atom-photon coupling $g_{i\ell}$ which depends on the photon ($i$) and atom ($\ell$) number. 
$a_i^\dagger$, $a_i$ are bosonic creation and annihilation operators, respectively, fulfilling canonical commutation relations.
$\sigma^{x,z}_{\ell}$ are spin-1/2 operators with Pauli matrix representation. As explained in detail in Refs.~\cite{dimer07,
gopa11}, the two states of the Ising spin in Eq.~(\ref{eq:dicke_hamiltonian}) map
onto two different stable ground-state sublevels, $|1\rangle$ and $|0\rangle$, of three-level $\Lambda$ atoms. 
$|1\rangle$ and $|0\rangle$ are indirectly coupled through a pair of Raman 
transitions to an excited state $|e\rangle$ which are driven by the classical field of a pair of external lasers. 
Upon adiabatic elimination of $|e\rangle$, one obtains Eq.~(\ref{eq:dicke_hamiltonian}) with 
$\sigma^z_\ell =|1_\ell\rangle\langle 1_\ell |- |0_\ell\rangle\langle 0_\ell | $ 
and $\sigma^x_\ell =|1_\ell\rangle\langle 0_\ell |+ |0_\ell\rangle\langle 1_\ell |$.
The parameters $\omega_i$, $\Delta$, and $g_{i\ell}$ can be controlled 
through laser frequencies and intensities. This tunability enables access to the strong-coupling 
Dicke regime. 
A dispersive shift of the cavity frequencies 
$\sim a^{\dagger}_i a_j\sigma^z$ does not modify our results significantly, and so will be set to zero for simplicity. 
A simple choice for a spatially varying atom-photon coupling is 
$g_{i\ell}=g \cos\left(k_i x_\ell\right)$ with $k_i$ the wavevector of the photon mode, and $x_\ell$ the coordinate 
of atom $\ell$.

In the single-mode, large photon wavelength case, we have $M=1$, $\omega_i = \omega_0$, 
and $g_{i \ell} = g/\sqrt{N}$ and the model can be solved exactly
in the $N\rightarrow\infty$ limit \cite{hepp73,wang73}. At zero temperature, 
there is a continuous phase transition between a paramagnetic phase (PM) and 
a superradiant ferromagnetic phase ($\text{FM}_{\text{SR}}$) at 
$g=g_{\text{c}}=\sqrt{\Delta\omega_0/8}$ at which the Ising symmetry 
$(a, \sigma_x)\rightarrow (-a,-\sigma_x)$, 
is spontaneously broken.

For the multimode Dicke model, it is useful to integrate out the photon degrees of freedom in 
a path-integral representation. Then the qubits are described by a Hamiltonian similar to the 
Ising model in a transverse field,
\begin{equation}
H_{\text{eff}} =
\frac{\Delta}{4}\sum_{\ell=1}^N\sigma_\ell^z
-
\frac{1}{2} \sum_{\ell m}J_{\ell m} \sigma^x_\ell \sigma^x_m\;,
\label{eq:dicke_eff}
\end{equation}
The exchange interactions $J_{\ell m}$ are mediated by the photons and have a frequency dependence associated with the photon 
frequencies $\omega_i$; thus Eq.~(\ref{eq:dicke_eff}) is to be understood as an action appearing in an imaginary time path-integral
summing over time-histories of the qubits. The long-range exchanges
\begin{equation}
J_{\ell m}(\Omega)=  \sum_{i=1}^M \frac{2 g_{i \ell} g_{i m} \omega_i}{\Omega^2+\omega^2_i}\;, 
\label{eq:J_freq}
\end{equation}
depend on $\Omega$, the imaginary frequency of the qubits in the path integral. 
Note that although we have formally integrated 
out the photons, we demonstrate below that the photon-photon correlation function is directly related to  
the atom-atom correlation function as obtained by solving Eq.~(\ref{eq:dicke_eff}).

If we ignore the frequency dependence in Eq.~(\ref{eq:J_freq}), the $J_{\ell m}$ have a structure similar to the Hopfield
model of associative memory \cite{amit85}, with $M$ `patterns' $g_{i \ell}$. For $M$ small, it is expected that such
a model can have $M$ possible superradiant ground states with $\text{FM}_{\text{SR}}$ order  $\left\langle \sigma^x_\ell \right\rangle
\propto g_{i\ell}$, $i=1 \ldots M$. In the spin-glass literature, these are the Mattis states which are ``good'' memories of 
the patterns $g$ \cite{amit85}.
The critical properties of the onset of any of these $\text{FM}_{\text{SR}}$ states should be similar
to those of the single mode Dicke model. 

Our interest in the present paper is focussed on larger values of $M$, where the summation in Eq.~(\ref{eq:J_freq}) can
be viewed as a sum over $M$ random numbers. 
Then, by the central limit theorem, the distribution of $J_{\ell m} ( \Omega)$ is Gaussian. 
Alternatively, the randomness of $J_{\ell m} (\Omega)$ can be enhanced by
passing the trapping laser beams through diffusers so that atomic positions $x_\ell$ are randomly distributed inside 
the cavity \cite{gopa11}. 
In either case, we assume a Gaussian distribution characterized by its mean and variance
\begin{eqnarray}
\overline{J_{\ell m} ( \Omega )} &=& J_0 (\Omega)/N 
\label{eq:prob} \\
\overline{\delta J_{\ell m} ( \Omega ) \delta J_{\ell' m'} (\Omega' )} &=& \left(\delta_{\ell \ell'} \delta_{m m'} + 
\delta_{m \ell'} \delta_{\ell m'} \right) K (\Omega, \Omega')/N, 
\nonumber
\end{eqnarray}
where the line represents a disorder average, and $\delta J_{\ell m}$ is the variation from the mean value.
We have assumed couplings between different sites are uncorrelated, and this will allow an exact solution in the $N\rightarrow \infty$ limit, 
modulo an innocuous softening of the
fixed length constraint on the Ising variable \cite{ye93,read95}. We will allow {\em arbitrary\/} frequency dependencies in $J_0 (\Omega)$
and $K(\Omega, \Omega')$. 
The factors of $N$ ensure an interesting $N \rightarrow \infty$ limit \cite{fischer91}. 
Especially for finite $M$, one could also use the methods of Ref.~\cite{amit85} to extend our analysis to models in which the 
$g_{i \ell}$ rather than the $J_{\ell m}(\Omega)$
are taken as independent random variables. However, as long as the photon modes can be chosen so that the
$J_{\ell m}(\Omega)$ vary in sign and magnitude, our analysis 
should remain qualitatively correct also for smaller values of $M$. 
%

%
%

\begin{figure}
\vspace*{1mm}
\includegraphics*[width=86mm,angle=0]{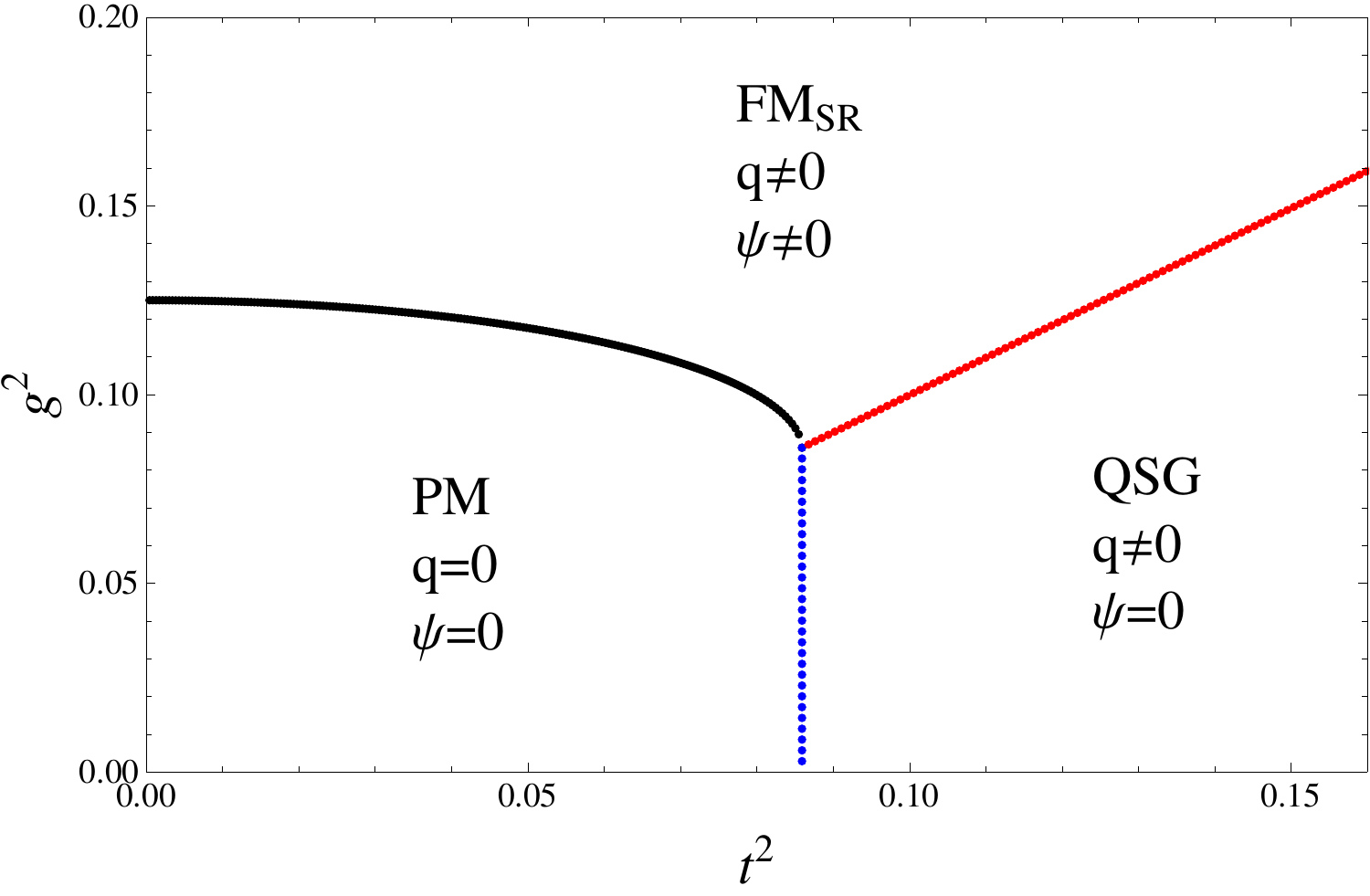}
\vspace*{-5mm}
\caption{(Color online) Zero-temperature phase diagram for $\omega_0=1$, $\Delta=1$ computed 
from Eq.~(\ref{eq:free_energy}). PM means 
paramagnet, $\text{FM}_{\text{SR}}$ superradiant ferromagnet, and QSG quantum spin glass. $q$ is the Edwards-Anderson 
order parameter and $\psi$ is the atomic population inversion or ferromagnetic order parameter.}
\label{fig:phasediag}
\end{figure}

{\textit{Key results}}.
We will show below that, in the limit of large atom number $N$, the results depend only upon
$J_0 ( \Omega = 0)$ and $K(\Omega, -\Omega)$. 
Here, we will display the phase diagram and spectral response functions for the simple 
choices $J_0 (0) = 2g^2/\omega_0$  and $K(\Omega, - \Omega) \equiv J^2 (\Omega)$ with
\begin{equation}
 J(\Omega)=2t^2\omega_0/(\Omega^2+\omega_0^2).
 \label{eq:choice_J}
\end{equation}
In Fig.~\ref{fig:phasediag}, we depict the ground 
state phase diagram; a related phase diagram in a condensed matter context was obtained in 
Ref.~\onlinecite{dali99}.
All phase transitions are continuous and the respective phase boundaries merge in a bicritical point at $(t^2_{\text{bc}}=0.086\,,\,g^2_{\text{bc}}=t^2_{\text{bc}})$. 
\begin{figure}
\vspace{1mm}
\includegraphics*[width=85mm,angle=0]{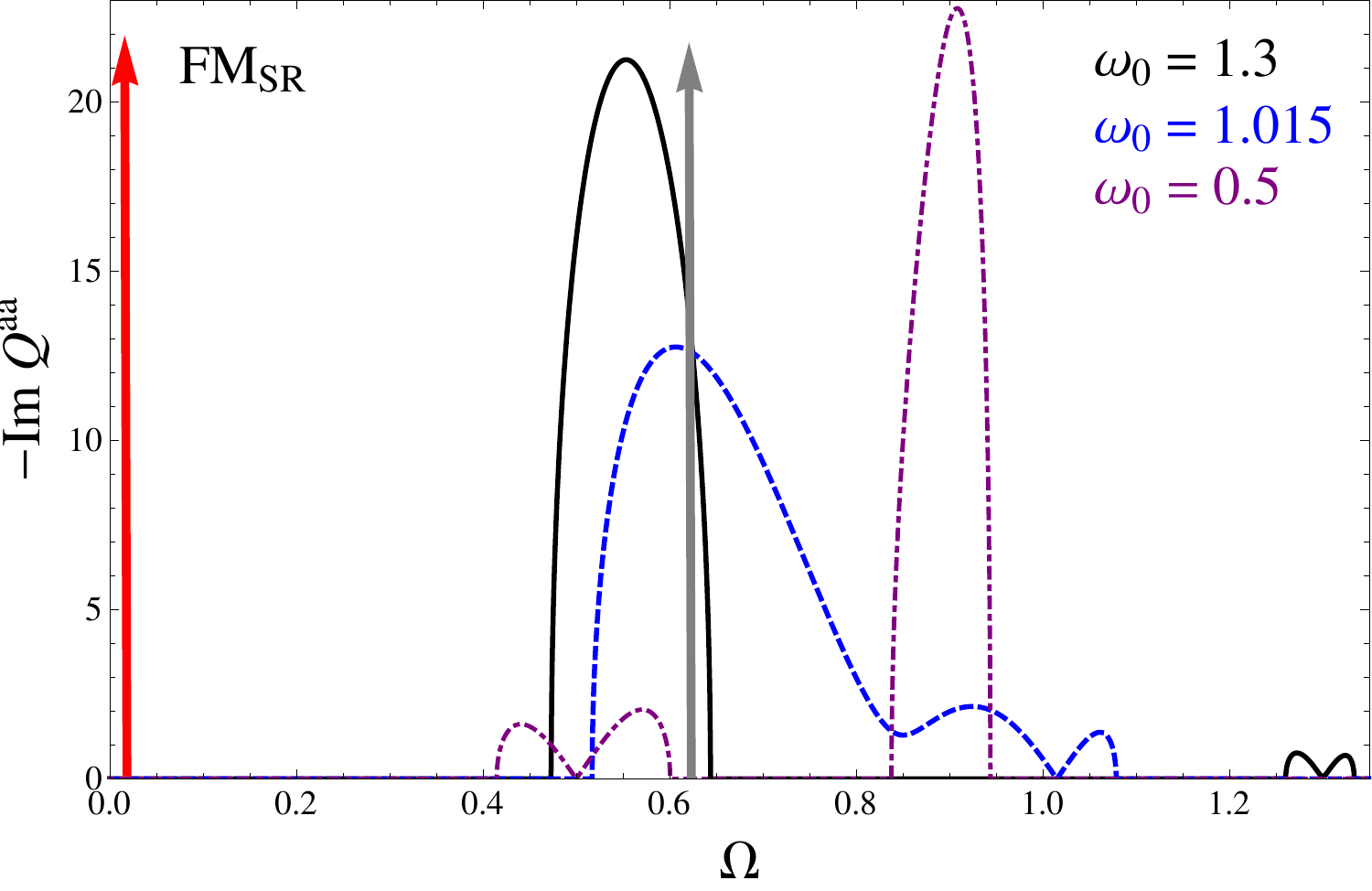}
\vspace*{0mm}
\caption{(Color online) rf spectral response function of the atomic qubits in the $\text{FM}_{\text{SR}}$ phase for various photon frequencies and $t^2=0.025$, $g^2=0.2$, $\Delta=1$. $\Omega$ is a real measurement frequency. The red arrow at $\Omega=0$ illustrates the delta function contribution with weight 
$q\sim\psi^2$ from Eqs.~(\ref{eq:spectral},\ref{eq:psi}). The value of the gap is given above Eq.~(\ref{eq:psi}). 
For the Dicke model without disorder ($t^2=0$), the spectral function following from 
Eq.~(\ref{eq:dicke_suscept}) consists of nothing but two delta functions: the red arrow at $\Omega=0$ and the grey arrow at 
$\Omega=g\sqrt{2\Delta/\omega_0}$ (plotted for $\omega_0=1.015$).}
\label{fig:spectral_fm}
\end{figure}

The intersection of the PM-$\text{FM}_{\text{SR}}$ phase boundary with the vertical axis at $t^2=0$ corresponds to the phase transition
in the single mode Dicke 
model without disorder \cite{emary03,dimer07}. In this case, a number of analytical results can be obtained from 
Eq. (\ref{eq:free_energy}), in agreement with the earlier work. The critical atom-photon coupling is $g_c^2=\Delta\omega_0/8$
 and the local $\sigma_\ell^x$ spin susceptibility in the $\text{FM}_{\text{SR}}$ phase is (for imaginary frequencies)
\begin{equation}
Q_{\ell}^{aa}(\Omega)\big |_{t^2=0}=\frac{\Delta}{\Omega^2+2\Delta g^2/\omega_0}+ \psi^2\,  2 \pi   \delta(\Omega).
\label{eq:dicke_suscept}
\end{equation}
The corresponding radiofrequency (rf) spectral response function of the atomic qubits for real frequencies, 
$-\text{Im}\left[Q^{aa}(i\Omega\rightarrow\Omega+i 0_+)\right]$, is depicted in Fig.~\ref{fig:spectral_fm}.
The superradiance is encoded in the zero frequency 
delta function contribution, whose weight is proportional to the atomic population inversion $\psi$. 
However, away from the zero frequency delta function, there is a spectral gap,
and the remaining spectral weight is a delta function at frequency $\sqrt{2 \Delta g^2/\omega_0}$.

The superradiance also appears as a photon condensate $\langle a_i \rangle = - \sum_{\ell} (g_{i \ell} /(2 \omega_i))
\langle \sigma^x_\ell \rangle$. We have computed the atomic population inversion, 
$\overline{\langle \sigma^x_{\ell} \rangle} = \psi$, and the Edwards-Anderson order parameter
$\overline{\langle \sigma^x_{\ell} \rangle^2} = q_{\text{QSG}}$ in Eqs. (\ref{eq:psi},\ref{eq:q_sg}). Both of these are related
to $\overline{\langle a_i \rangle}$, but computation of the latter requires more specific knowledge of the $g_{i \ell}$.
For $\Omega \neq 0$, the photon correlation function follows from Eq.~(\ref{eq:dicke_suscept}) 
\begin{equation}
\langle a_i^{\dagger}(\Omega)a_j (\Omega)\rangle=\left[\left(i\Omega-\omega_i\right)\delta_{ij}
+\sum_{\ell=1}^N g_{i\ell}g_{j\ell}Q_\ell^{aa}(\Omega)\right]^{-1}\;,
\label{eq:photon_spec}
\end{equation}
where the right-hand-side is a matrix inverse, as can be obtained from integrating out the atomic 
fields from the path-integral representation of Eq.~(\ref{eq:dicke_hamiltonian}).

Upon introducing small disorder (with $t\neq0$), as long as we remain in the $\text{FM}_{\text{SR}}$ phase, the zero frequency delta
function and spectral gap survive, although the higher frequency spectral weight changes, as shown in Fig.~\ref{fig:spectral_fm}.
This spectral gap is present across the phase transition from the $\text{FM}_{\text{SR}}$ phase to the PM phase. Thus all the low energy
fluctuations in the critical theory for this transition are restricted to the zero frequency delta function, which can be described in classical theory for
the spins: this is the reason this transition is more properly considered as a {\em classical\/} phase transition.

\begin{figure}
\includegraphics*[width=85mm,angle=0]{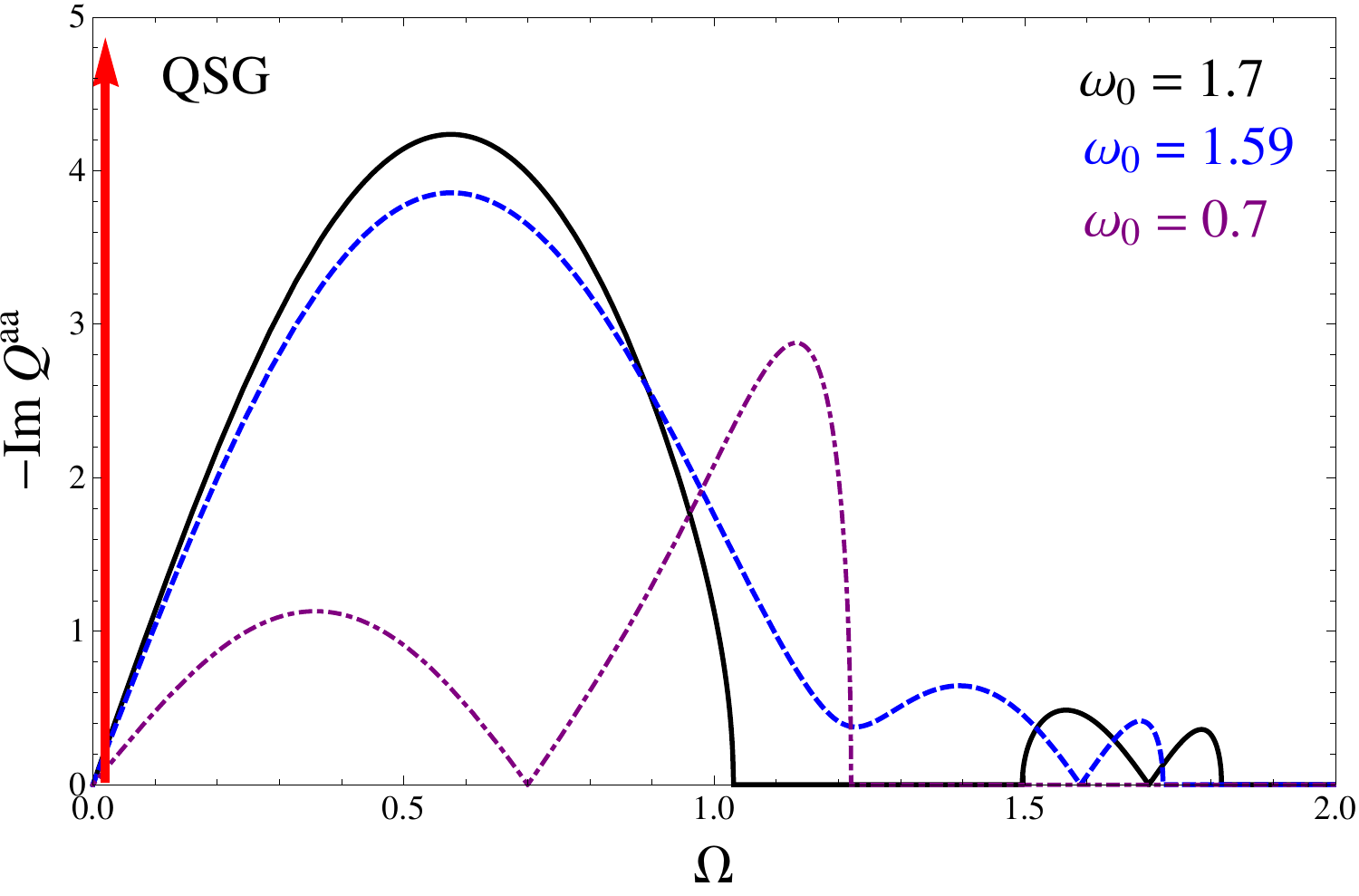}
\vspace*{0mm}
\caption{(Color online) rf spectral response function of the atomic qubits in the QSG phase for various photon frequencies and $t^2=0.175$, $g^2=0.05$, $\Delta=1$. 
The red arrow at $\Omega=0$ illustrates the delta function contribution with weight $q\sim q_{\text{QSG}}$ from 
Eqs.~(\ref{eq:spectral},\ref{eq:q_sg}).}
\label{fig:spectral_sg}
\end{figure}
For a sufficiently large value of $t^2$, the system undergoes 
a quantum phase transition to the QSG ground state. 
In contrast to the PM-$\text{FM}_{\text{SR}}$ transition, at the QSG transition, and in the entire QSG phase, there is spectral
weight at a continuum of frequencies reaching zero
(see Fig.~\ref{fig:spectral_sg}). Thus the onset of QSG order from the PM phase is a genuine {\em quantum\/} phase transition, whose
universality class was described in Ref.~\onlinecite{read95}.

The PM phase is clearly delineated from both, the QSG and the $\text{FM}_{\text{SR}}$ phases: the PM phase 
has a gapped spectral response and no superradiant photon condensates. 

We also note that in all phases, while the spectral function has a universal form at low frequencies, 
its high frequency behavior is strongly dependent upon the forms of $J_0 (\Omega)$ and $J(\Omega)$.
For the forms in Eq.~(\ref{eq:choice_J}), the spectral function is suppressed to zero at
 $\Omega=\omega_0$.
 
 {\textit{Experimental signatures}}.
The rf spectral response function of the atomic qubits presented in Figs.~\ref{fig:spectral_fm},\ref{fig:spectral_sg}
should be observable via radiofrequency spectroscopy \cite{stewart08,haussmann09}.

Measuring the spectrum of photons leaving the cavity through its imperfect mirrors at loss rate $\kappa$ 
allows for an in-situ measurement of our phase diagram, Fig.~\ref{fig:phasediag}. 
Our prediction for the spectrum of intra-cavity photons, Eq.~(\ref{eq:photon_spec}), can be 
related to the extra-cavity photons via the input-output formalism 
\cite{dimer07,gardiner84, walls08}. For this case of a dissipative Dicke model, 
we note a similarity of the decay effects to those in theories of {\em metallic\/} spin glasses \cite{sro}, 
in which the spin qubits are coupled to a ``reservoir'' of continuum spin excitations near the Fermi surface. 
This coupling leads to a damping term
in the dynamics of each spin, but does not significantly modify the spin-spin interactions responsible for the 
spin glass phase. Similarly, for the dissipative Dicke model, decay into photons outside the cavity will
introduce various damping terms {\em e.g.\/} a $\kappa |\Omega|$ term in the denominator of 
Eq.~(\ref{eq:J_freq}). As in the previous analyses \cite{sro}, 
we expect that the quantum spin glass transition will survive in the presence of damping, although there will be 
some changes to the critical properties \cite{follow-up}.

As in other glasses, we expect slow relaxational dynamics, along with memory and aging effects in the QSG phase 
which should be observable via local spin addressing protocols and measuring the spin relaxation time 
scale \cite{gopa11}. 

{\textit{Conclusion}}.
Observations of these effects in quantum optic systems would be remarkable. Moreover, 
the spin glass physics is driven by long-range interactions, and this makes 
the theoretical models analytically tractable. 
We therefore have prospects for a quantitative confrontation between theory 
and experiment in a glassy regime, something which has eluded other experimental realizations of 
spin glasses.

%
%
{\textit{Details of the calculation}}. 
As discussed in Refs.~\onlinecite{ye93,read95}, each Ising qubit, with on-site gap $\Delta/2$, is conveniently represented
by fluctuations of a non-linear oscillator $\phi_\ell (\tau)$ ($\tau$ is imaginary time) which obeys a unit-length constraint.
Their action at temperature $T$ is then
\begin{equation}
\mathcal{S}_0 [ \phi, \lambda ] = \frac{1}{2\Delta} \sum_{\ell = 1}^N \int_0^{1/T} d \tau \left[
(\partial_\tau \phi_\ell)^2 + i \lambda_\ell ( \phi_\ell^2 - 1 ) \right]
\end{equation}
where $\tau$ is imaginary time, and the $\lambda_\ell$ are Lagrange multipliers which impose the constraints.
The {\em only\/} approximation of this paper is to replace the $\lambda_\ell$ by their saddle-point value, $i \lambda_\ell = \lambda$, and 
to ignore their fluctuations.
For decoupled oscillators, this saddle-point value is $\lambda=\Delta^2/4$, the $\phi$ susceptibility is $\Delta/(\Omega^2 + \Delta^2/4)$, 
and the resulting gap, $\Delta/2$, has been matched to that of the Ising spin.

The interactions between the qubits are accounted for as before \cite{ye93}: we introduce replicas $a=1 \ldots n$, average over
the $J_{\ell m}$ using Eq.~(\ref{eq:prob}), decouple the resulting two-$\phi$ coupling by Hubbard-Stratonovich transformation using a ferromagnetic order parameter $\Psi^a (\Omega)$, and the four-$\phi$ coupling by the bilocal field 
$Q^{ab} (\Omega_1, \Omega_2)$ \cite{read95} (the $\Omega$ are Matsubara frequencies).
The complete action is
\begin{eqnarray}
&& \mathcal{S} = \sum_a \mathcal{S}_0 [\phi^a, \lambda^a] + T \sum_{a,\Omega} J_0 (\Omega) \Biggl[ \frac{N}{2} |\Psi^a (\Omega)|^2 \nonumber \\ && -   \Psi^a (-\Omega) \sum_{\ell=1}^{N} \phi_\ell^a (\Omega) \Biggr]  + \frac{T^2}{2} \sum_{a,b,\Omega,\Omega'}
\!\!\!\! K (\Omega, \Omega' ) \Biggl[ \frac{N}{2} |Q^{ab} (\Omega, \Omega') |^2 \nonumber \\
&&~~~~~~~~ -  Q^{ab} (- \Omega, 
- \Omega' ) \sum_{\ell=1}^N \phi^a_\ell (\Omega) \phi_\ell^b (\Omega') \Biggr]\;. \label{action}
\end{eqnarray}
Now we perform the Gaussian integral over the $\phi_\ell$: the resulting action has a prefactor of $N$, and so can be replaced
by its saddle-point value. By time-translational invariance, the saddle-point values of the fields can only have the
following frequency dependence
\begin{eqnarray}
\Psi^a ( \Omega) &=& \left( \delta_{\Omega, 0}/T \right) \psi \nonumber \\
Q^{ab} (\Omega, \Omega' ) &=& \left( \delta_{\Omega + \Omega',0}/T \right) \left[ \chi (\Omega) \delta^{ab} + 
\left( \delta_{\Omega,0}/T \right) q  \right], \label{ansatz}
\end{eqnarray}
and we take $\lambda^a = \lambda$. 
We have assumed replica symmetry for the Edwards-Anderson order parameter $q$ because our interest will be limited here
to $T=0$ where there is no replica symmetry breaking \cite{read95}.
Now the values of the ferromagnetic moment $\psi$, the 
spin susceptibility $\chi ( \Omega)$, $q$, and $\lambda$ have to be determined
by optimizing the free energy. The latter is obtained by inserting Eq.~(\ref{ansatz}) in Eq.~(\ref{action});
after taking the replica limit $n \rightarrow 0$, we have the free energy per site
\begin{eqnarray}
&& \mathcal{F} =  \frac{J_0 (0)  \psi^2}{2} + \frac{T}{4} \sum_\Omega K(\Omega, -\Omega) | \chi ( \Omega ) |^2  + 
\frac{1}{2}K(0,0) \chi (0) q \nonumber \\
&&~~~~~~ + 
\frac{T}{2} \sum_\Omega \ln \left( \frac{(\Omega^2 + \lambda)}{\Delta} - K(\Omega,-\Omega) \chi (\Omega) \right) 
 - \frac{\lambda}{2 \Delta}
\nonumber \\
&&~~~~~~   - \frac{1}{2} \left[\frac{K(0,0) q + J_0^2 (0) \psi^2}{\lambda/\Delta - K(0,0) \chi (0) }\right]\;.
\label{eq:free_energy}
\end{eqnarray}
Note that this free energy depends only upon $J_0 (0)$ and $K(\Omega, - \Omega)$, as claimed earlier. 
Our results described in Eq.~(\ref{eq:dicke_suscept}) and Figs.~\ref{fig:phasediag}-\ref{fig:spectral_sg} are derived from a set 
of coupled saddle-point equations obtained from varying Eq.~(\ref{eq:free_energy}) with respect to 
$\chi(\Omega)$, $q$, $\psi$, and $\lambda$ for every $\Omega$. Subsequently we let $T\rightarrow 0$.

%
%

For the choices for $K(\Omega,-\Omega)$ and $J_0(0)$ of Eq.~(\ref{eq:choice_J}), 
the rf spectral response function of the atomic qubits plotted in figures \ref{fig:spectral_fm},\ref{fig:spectral_sg} 
is given by the expression:
\begin{align}
&-\text{Im}\left[Q^{aa}(i\Omega\rightarrow\Omega+i 0_+)\right]=\label{eq:spectral}\\
&\frac{\left|\omega_0 ^2-\Omega ^2\right| \sqrt{16 \Delta ^2 t^4 \omega_0^2-\left(\lambda -\Omega ^2\right)^2
   \left(\omega_0^2-\Omega ^2\right)^2}}{8 \Delta  t^4 \omega_0^2}+q\, 2\pi \delta(\Omega)\;.\nonumber
\end{align}
The first term is non-zero only for frequencies $\Omega$ so that the expression underneath the square-root is 
positive. The value of the Lagrange multiplier in the $\text{FM}_{\text{SR}}$ is pinned to
$\lambda_{\text{FM}}=\Delta\left(J_0(0)+{K(0,0)}/{J_0(0)}\right)$. The value of the 
gap in Fig.~\ref{fig:spectral_fm} is 
$
\sqrt{\frac{1}{2}\left(\lambda_{\text{FM}}+\omega_0^2 -\sqrt{16 \Delta  t^2 \omega_0+
\left(\lambda_{\text{FM}} -\omega_0 ^2\right)^2}\right)}\;.
$
This expression equates to zero in the gapless QSG phase shown in Fig.~\ref{fig:spectral_sg}, 
where $\lambda_{\text{QSG}}=2\Delta\sqrt{K(0,0)}$. 
This gap vanishes logarithmically faster than $(t^2-t^2_c)$ when approaching the QSG phase boundary due to the 
square-root behavior of the spectral weight \cite{ye93,miller93}.

The ferromagnetic moment obtains as 
\begin{equation}
\psi^2=\frac{J^2_0(0)-K(0,0)}{J^2_0(0)}    
\left(1-\int^{\infty}_{-\infty}\frac{d\Omega}{2\pi}\chi(\Omega)
\Big |_{\lambda=\lambda_{\text{FM}} }\right) \label{eq:psi}\;,
\end{equation}
and $\psi$ vanishes continuously at the $\text{FM}_{\text{SR}}$-QSG phase boundary
(at which $J_0(0)=\sqrt{K(0,0)}$) with exponent $\beta_{\text{FM}}=0.5$. As per the 
discussion above Eq.~(\ref{eq:photon_spec}), the corresponding photon condensate
$\langle a_i \rangle$ vanishes with the same exponent. 
Note that the Edwards-Anderson order parameter $q$ is continuous across this transition 
and in the QSG phase given by:
\begin{equation}
q_{\text{QSG}}=1-\int^{\infty}_{-\infty}\frac{d\Omega}{2\pi}\chi(\Omega)
\Big |_{\lambda=\lambda_{\text{QSG}} }\;.
\label{eq:q_sg}
\end{equation}
As expected, one obtains numerically $\beta_{\text{QSG}}=1.0=2\beta_{\text{FM}}$. 

{\textit{Acknowledgments}}. 
We thank A.~Amir, J.~Bhaseen, T.~Esslinger, J.~Keeling, B.~Lev, M.~Punk, J.~Ye, P.~Zoller, and especially J.~Simon for 
useful discussions. This research was supported by the DFG under grant Str 1176/1-1, 
by the NSF under Grant DMR-1103860, and by a MURI grant from AFOSR.

\end{document}